\pdfoutput=1

\documentclass{JASSS}

\doinum{10.18564/jasss.4252}
\volume{23}
\issue{2}
\article{6}
\pubyear{2020}
\received{17-07-2019}
\accepted{05-02-2020}
\published{31-03-2020}

\title{Homophily as a process generating social networks:
insights from Social Distance Attachment model}

\reviewcopy{false}

\author[1]{Szymon Talaga}
\author[1,2]{Andrzej Nowak}
\affil[1]{The Robert Zajonc Institute for Social Studies, University of Warsaw}
\affil[2]{Department of Psychology, University of Warsaw}

\email{stalaga@protonmail.com}

\usepackage{natbib}
    \setcitestyle{authoryear,round,aysep={}}

\usepackage[utf8]{inputenc} 
\usepackage[T1]{fontenc}    
\usepackage{amsmath}
\usepackage{amsfonts}       
\usepackage{nameref}        
\usepackage[
    colorlinks = true,
    citecolor = blue,
    linkcolor = blue
]{hyperref}

\newcommand{\Space}{\mathcal{S}}
\newcommand{\R}{\mathbb{R}}
\newcommand{\E}{\mathbb{E}}
\newcommand{\mat}[1]{\boldsymbol{#1}}

\graphicspath{{./figures/}}

\begin{document}
\maketitle



\begin{abstract}
    Real-world social networks often exhibit high levels of clustering,
    positive degree assortativity, short average path lengths (small-world property)
    and right-skewed but rarely power law degree distributions.
    On~the other hand homophily, defined as the propensity of similar agents
    to connect to each other, is one of the most fundamental social processes
    observed in many human and animal societies.
    In this paper we examine the extent
    to which homophily is sufficient to produce the typical structural
    properties of social networks. To do so, we conduct a simulation study
    based on the Social Distance Attachment (SDA) model, a particular kind of
    Random Geometric Graph (RGG), in which nodes are embedded in a social
    space and connection probabilities depend functionally on distances between
    nodes. We derive the form of the model from first principles based on
    existing analytical results and argue that the mathematical construction
    of RGGs corresponds directly to the homophily principle,
    so they provide a good model for it.
    We find that homophily, especially when combined with a random edge rewiring,
    is sufficient to reproduce many of the characteristic features of social networks.
    Additionally, we devise a hybrid model combining SDA with the configuration
    model that allows generating homophilic networks with arbitrary
    degree sequences and we use it to study interactions of homophily
    with processes imposing constraints on degree distributions.
    We show that the effects of homophily on clustering
    are robust with respect to distribution constraints, while degree
    assortativity can be highly dependent on the particular kind of
    enforced degree sequence.
\end{abstract}


\parano{}













\section{Introduction}\label{sec:intro}

Networks provide one of the most useful analytical and theoretical frameworks for
studying social phenomena. Real-world social networks often exhibit a set
of characteristic structural properties that do not occur jointly as often
in different kinds of complex networks
(i.e. technological, informational, biological).
In particular, they tend to be sparse, have non-trivial clustering coefficients,
positive degree assortativity,
and short average path lengths (small-world property).
Moreover, social networks often have right-skewed
but rarely power law degree distributions
\citep{newman_why_2003,watts_collective_1998,boguna_models_2004,broido_scale-free_2019}.
Therefore, one should account for these characteristic properties when
modeling social networks. This is especially important in the context
of simulation studies, in which researches often either explicitly
or implicitly assume some underlying network structure,
but sometimes lack appropriate tools to generate or model such structures
and have to resort to well-known generic models such as Erdős-Rényi random graphs
\citep{erdos_random_1959}
or preferential attachment networks
\citep{barabasi_emergence_1999}
that may not fit well to the social problem
at hand. Moreover, proper understanding of how various social processes affect
networks' structures and \textit{vice versa} is also theoretically
and practically important for many fields and problems within social sciences
such as opinion dynamics, diffusion of information/diseases
\citep{flache_models_2017,sobkowicz_modelling_2009,stroud_spatial_2007},
consensus emergence
\citep{stocker_consensus_2001},
social impact theory and dynamical social psychology
\citep{latane_psychology_1981,nowak_private_1990,nowak_dynamical_1998},
as well as many kinds of agent-based models
\citep{bianchi_agent-based_2015}.
Perhaps even more importantly, this question
constitutes also the central problem in sociological theories of social structure.

Social structure and social networks are related through the fundamental
principle of homophily, according to which agents that are similar to each other
with respect to some significant social features are more likely to be somehow
connected than dissimilar agents
\citep{blau_macrosociological_1977,mcpherson_ecology_1983,popielarz_edge_1995,mcpherson_birds_2001}.
Social structure corresponds to a distribution of social positions
(combinations of features), and relative positions determine
probability of an occurrence of a tie of a given type between any two agents.
This idea is reflected in sociology in a long and rich tradition of thinking
about social structure in terms of social spaces\footnote{
    Social spaces in the works of Blau and Bourdieu are two quite different concepts,
    but they are both formulated in explicit geometric terms.
}
in which actors are represented as points and relations
between them follow from distances in the space
\citep{blau_macrosociological_1977,mcpherson_blau_2004,bourdieu_distinction_1986,bourdieu_social_1989}.
In other words, structure of social networks is directly linked to geometry
(distribution of positions) of social spaces they are embedded in.

Thus, it is quite natural to represent social networks with various models
of spatially-embedded graphs such as
Random Dot Product Graphs
\citep{scheinerman_modeling_2010},
Latent Space Model
\citep{hoff_latent_2002}
or Social Circles
\citep{hamill_social_2009}.
However, perhaps one of the most generic and parsimonious models
are Random Geometric Graphs (RGG),
in which connection probabilities are modeled explicitly with non-increasing
functions of distances in an embedding space
\citep{dall_random_2002,krioukov_clustering_2016},
since their mathematical formulation corresponds
directly to the sociological notions of social structure and homophily.

The aim of this paper is to examine the extent to which the principle of homophily
is sufficient to explain the typical structural features of social networks
and to determine whether it can be the sole generating process or rather needs
to be complemented by other processes. This question is important
because the link between structure of social networks and homophily is already
a well-established sociological fact
\citep{blau_macrosociological_1977,mcpherson_birds_2001}
and homophily itself is one of the most universal processes observed
in social systems
\citep{mcpherson_homophily_1987,marsden_core_1987,kossinets_origins_2009},
socio-technological systems
\citep{aiello_friendship_2012,colleoni_echo_2014,halberstam_homophily_2016}
and even animal societies
\citep{lusseau_identifying_2004,jiang_assortative_2013}.
This is not a new question of course, as already a great deal of work on this
problem has been done
\citep[e.g.][]{popielarz_edge_1995,centola_homophily_2007,carletti_emerging_2011,centola_choosing_2015}.
However, the novelty of our contribution follows from
the fact that we use a very general formal model of
homophily-driven network formation which we derive from first principles.
Moreover, we explicitly address the question of the extent to which homophily
is sufficient to reproduce structural properties considered typical for
social networks. We also examine its robustness with respect to several factors
such as system size changes and combining with other processes.

To achieve our goal we first use existing analytical results to show
an approximate equivalence between social networks shaped by homophily and
random geometric graphs and derive the proper form of the connection
probability function. Next we use and extend a particular type of RGG
introduced by
\citet{boguna_models_2004},
which we call the Social Distance Attachment (SDA) model,
to conduct simulation studies to validate our analysis and extend it beyond
the existing analytical results. We run SDA simulations for much wider range
of parameters than the one used by the authors in the original paper
\citep{boguna_models_2004}
and extend the model so one can control expected average degree
as well as enforce arbitrary degree sequences. The goal of our simulation study
is to assess the extent to which pure homophily or homophily combined with other
processes such as a random edge rewiring and degree distribution constraints
can reproduce the typical properties of social networks.
Through all this we want both to provide a theoretical insight into the
role homophily may play as a process structuring social networks,
as well as equip researchers with graph models that are both simple,
interpretable and useful for generating artificial networks
that reproduce some of the most typical features of real-world social networks.

The rest of the paper is structured as follows. First, we briefly define
network metrics that will be used later and discuss in more detail structural
properties that are typical for social networks.
Then, we position our approach to the problem within the broader literature
about homophily and social networks.
Next, we discuss the connection between homophily and random geometric graphs
based on existing analytical results.
After that we introduce the SDA model and discuss the results of the first
simulation study.
Then, we introduce a hybrid of SDA and the configuration model that allows
generating homophilic networks with arbitrary degree sequences
and discuss the results of the second simulation study.
Last but not least, we discuss all the results in the light of the existing
literature on social networks and homophily and use them to propose a crude,
but easy-to-apply, rule of thumb that may help to identify real-world
social networks which may be shaped by homophily to a~significant degree.

\subsection{Code and replication}

Python implementation of both models and scripts replicating the simulations
as well as R code for the data analysis are
available at GitHub:
\href{https://github.com/sztal/sda-model}{https://github.com/sztal/sda-model}.
The repository contains both code and documentation.
Frozen version of the repository can be accessed through
\textit{CoMSES} library\footnote{
    \href{https://www.comses.net/codebases/1adc39e5-344c-428c-9a20-fd53d791709b/releases/1.1.0/}{https://www.comses.net/codebases/1adc39e5-344c-428c-9a20-fd53d791709b/releases/1.1.0/}
}.

\section{Properties of social networks}\label{sec:properties}

Here we define network metrics which we will use and
describe in detail the characteristic features of social networks.
We focus on the following structural properties:

\begin{itemize*}
    \item \textbf{Sparsity.} A graph is sparse if its mean degree ($\E[k]$) grows
    sublinearly with the system size/number of nodes ($N$). This implies
    that edge density (fraction of existing edges) goes to zero with increasing
    $N$. In particular, networks with fixed mean degree are sparse
    \citep{newman_networks:_2010}. Sparsity is typical for many types of
    real-world networks, but it is especially important for social networks.
    Fixed average degree in social networks is implied by the notion of Dunbar's
    numbers, which corresponds to the typical amount of meaningful relationships
    (of a given closeness) a person may effectively maintain
    \citep{hill_social_2003,mac_carron_calling_2016}.
    Hence, any model of social networks, at least in the case of relations
    between individuals, should take this into account.

    \item \textbf{Non-trivial clustering.} Clustering (transitivity),
    measures probability that if a randomly selected node is connected
    to two other nodes, then these two nodes are also connected
    (such property is called triadic closure).
    In the case of dense graphs clustering may be trivial.
    Consider an Erdős-Rényi random graph
    \citep{erdos_random_1959}
    in which every possible edge occurs with probability $p$
    (hence $\E[k] = p(N-1)$ grows linearly with $N$).
    In such a graph expected clustering is $p$
    \citep[p.~347]{newman_networks:_2010}
    and is equal to the edge density. In sparse networks density goes to zero,
    so if there is non-zero clustering, even for large $N$, then there must
    be some non-trivial process that enforces triadic closure. This is exactly
    why joint sparsity and high clustering is a very important structural
    feature. It is considered one of the hallmarks of social networks
    \citep[p.~200]{newman_why_2003,newman_networks:_2010}.
    If not stated otherwise, by clustering we always refer to the global
    clustering / transitivity, rather than the mean nodal clustering coefficient
    \citep[p.~198--204]{newman_networks:_2010}.

    \item \textbf{Positive degree assortativity.} Degree assortativity
    quantifies the tendency of nodes to connect to other nodes of similar degree.
    It is defined as a Pearson correlation between degrees of adjacent nodes
    \citep{newman_assortative_2002}.
    Not all social networks have positive degree assortativity ---
    for instance the famous Zachary's karate club network is disassortative.
    However, many do and positive assortativity is generally considered typical
    for a broad class of social networks
    \citep{newman_why_2003}.

    \item \textbf{Small-world property.} A network is a small-world if its
    average shortest path ($L$) between nodes grows proportionally to $\log N$
    \citep{watts_collective_1998}.
    The property can be formally assessed only on the basis of a generative
    model defining scaling of average path lengths with respect to the growth
    of the network size, but empirical studies showed that in large real-world
    social networks average shortest paths tend to be around 6
    \citep{travers_experimental_1969,leskovec_planetary-scale_2008}.
    In other words, average shortest paths in social networks tend to remain
    short even when a network grows very large.

    \item \textbf{Diverse but bounded degree distribution.} A degree distribution
    can be considered bounded if at least its first two moments
    (expected value and variance) are finite,
    so the Central Limit Theorem applies.
    Most typically unbounded degree distributions are power laws.
    Of course researchers in social sciences
    usually work with observed networks which have fixed size and finite
    moments by definition and not so much with analytical models
    and asymptotic methods, so it is perhaps natural to expect
    degree distributions of social networks
    to be bounded. However, boundedness is rather a property of
    the data-generating process than the data itself, so it may be still
    worthwhile to consider the question of the prevalence of power laws
    in degree distributions of social networks. In fact, there are good reasons
    to expect many social networks to have bounded (not power law)
    degree distributions as this is implied by the Dunbar's numbers
    and at least partially supported by data
    \citep{broido_scale-free_2019}.
    Hence, we argue that the boundedness of degree distributions should be included
    in the list of typical properties of social networks. However, it~does not
    imply that degree distributions in social networks are exclusively
    Poisson-like and symmetric. Quite the contrary, they may come in many
    different shapes, in particular markedly right-skewed ones
    \citep{newman_why_2003,boguna_models_2004,broido_scale-free_2019}.
\end{itemize*}

It should be stressed that many other types of complex networks
(i.e. technological, informational or biological) may exhibit some of the
above-mentioned properties, but will usually also differ in some important
respects. For instance, power laws are more common at least
among biological and technological networks
\citep{broido_scale-free_2019},
clustering levels in non-social networks are more often very close
to what is induced just by degree distributions
(this may be considered a type of trivial clustering)
\citep{newman_why_2003},
and food webs are usually dense
\citep[p.~135]{newman_networks:_2010}.

\subsection{Homophily and geometry of social networks}\label{sec:geometry}

It is important to note that there are several different kinds of homophily.
Usually in the literature one can find a~distinction between \textit{value}
and \textit{status} or \textit{choice} and \textit{induced} homophily
\citep{mcpherson_homophily_1987,mcpherson_birds_2001}. Value/choice homophily
refers to similarity between agents that is due entirely to their more or less
conscious preferences. For instance highly extrovert people may prefer
to spend time with others who are also extrovert. On the other
hand status/induced homophily refers to similarity that follows from structural
constraints imposed on social agents such as geographical distance, racial
and ethnic segregation~etc.

Moreover, from the dynamical perspective similarity between connected social
agents may result not only from selection (homophily), but also from influence,
as when two individuals are initially dissimilar,
but with time become more alike through repeated social interactions
\citep{flache_models_2017}. This indicates that similarity in social networks
is linked to two distinct causal mechanisms. It is either that similarity
breeds connection or connection breeds similarity, and it is usually very
difficult to disentangle these two effects
\citep{anagnostopoulos_influence_2008}.

However, from a static point of view the problem of selection and influence
disappears, since these are dynamic processes. When looking
at a single snapshot of a social network any statement about homophily is
a correlational statement as it refers to a probabilistic pattern that
any two connected agents are on average more similar than any two disconnected
agents, irrespective of any causal/dynamic processes that might have produced it.
Moreover, in this case the distinction between value/choice and
status/induced homophily also reduces just to the issue of interpretation
of dimensions of a social space. Therefore, the general problem of the connection
between homophily and structure of social networks naturally decomposes into
two parts:

\begin{enumerate*}
    \item \textbf{Static part.} What would be the typical structure of a social
    network assuming that agents who are more similar to each other
    (with respect to a set of features) are more likely to connect?
    \item \textbf{Dynamic part.} Which structural features of social networks
    can be attributed to selection
    (subsequently decomposed into value/choice and status/induced homophily)
    and which to influence and to what extent?
\end{enumerate*}

In this paper we try to address the first part. To do so, we first note that
random geometric graph (RGG) models, in which
all connection probabilities are derived simultaneously from distances
in an embedding space according to a fixed formula
\citep{dall_random_2002,krioukov_clustering_2016},
provide a very convenient framework for this problem, since their
mathematical formulation corresponds exactly to the structure of the social
process. Thus, now we turn to a brief overview of the main properties
of RGG in order to further explicate their important connection to social networks
and homophily.

First, we show how RGG emerge naturally in the context of social networks
based on very weak and justifiable assumptions.
We consider a family of networks in which
expected node degree and node-level clustering are fixed to some positive values.
It has been proved that under such assumptions the maximum entropy distribution
over the family of networks meeting the constraints is a RGG model with nodes
distributed uniformly along the real line and the connection probability
function approximated by the Fermi-Dirac distribution\footnote{
    Fermi-Dirac distribution is used in quantum physics and is usually
    defined with a slightly different but equivalent parametrization.
    Here we present it in a form that is most easily interpreted in terms
    of homophily. The larger system size, the better the approximation.
}
\citep{krioukov_clustering_2016}:
\begin{equation}\label{eq:fd}
    p_{ij} = \frac{1}{1 + e^{\alpha(d(\mat{x}_i, \mat{x}_j) - b)}}
\end{equation}
where $p_{ij}$ is a connection probability for nodes $i$ and $j$,
$d(\mat{x}_i, \mat{x}_j)$ is a distance between the nodes in an embedding space,
$b$ is the characteristic distance at which $p_{ij} = 1/2$
and $\alpha$ is the homophily parameter that controls how fast $p_{ij}$
goes to zero when the distance goes to infinity and how fast it goes towards $1$
when the distance approaches $0$. Note that this is a sigmoidal
function and the edge density (average $p_{ij}$) is controlled jointly by
$b$ and $\alpha$. However, once $\alpha$ is fixed, density depends only on $b$,
so it can be used to set a fixed average degree regardless of the system size
(at least when $\alpha$ is high enough). Hence, the model allows for sparsity.

The fact that the obtained probability distribution is maximum entropy implies
that among all possible probability distributions meeting the constraints
this one is the most unbiased with respect to any network properties
other than the expected degree and local clustering.
Thus, sparsity and clustering (two of the typical features of social networks)
leads naturally to RGG.

Now let us note that in the limit of large homophily ($\alpha \to \infty$)
the connection probability function reduces to a step function:
\begin{equation}\label{eq:rgg}
    p_{ij} =
        \begin{cases}
            1,& \text{if } d(\mat{x}_i, \mat{x}_j) < b\\
            0,& \text{if } d(\mat{x}_i, \mat{x}_j) > b
        \end{cases}
\end{equation}
This is the so-called hard RGG model
\citep{dall_random_2002}.
In fact the reduction will happen for any connection probability function
that is sigmoidal and parametrized by homophily ($\alpha$) and characteristic
distance ($b$).
It is useful because it has been shown that at least for
spaces with uniform density hard RGGs have clustering that does not vanish with
increasing system size, even when edge density asymptotically goes to zero
(sparsity). Moreover, hard RGGs also exhibit positive degree assortativity
\citep{antonioni_degree_2012}.
This means that (hard) RGGs naturally lead to graphs with several typical
properties of social networks.

Combining the two analytical results together we see that there is a kind of an
approximate equivalence between networks with the first three typical properties
of social networks we listed and random geometric graphs,
since we have a biconditional:
\begin{itemize*}
    \itemsep0.5em
    \item
    \begin{minipage}{.3\textwidth}
        \textbf{Sparsity and clustering}
    \end{minipage}
    \begin{minipage}{.1\textwidth}
        $\Longrightarrow$
    \end{minipage}
    \begin{minipage}{.5\textwidth}
        \textbf{RGG}
    \end{minipage}
    \item
    \begin{minipage}{.3\textwidth}
        \textbf{RGG}
    \end{minipage}
    \begin{minipage}{.1\textwidth}
        $\Longrightarrow$
    \end{minipage}
    \begin{minipage}{.5\textwidth}
        \textbf{sparsity, clustering and positive degree assortativity}
    \end{minipage}
\end{itemize*}
The equivalence is exact only in the limit of large system size and high
homophily, but it strongly suggests that there is an important connection
between a particular class of social networks
(sparse ones with high clustering and positive degree assortativity)
and random geometric graphs. This confirms the idea that RGGs provide a proper
framework for studying the connections between homophily and structure
of social networks, or at least the first (static) part of this problem in our
formulation. However, the analytical results we reviewed rely on the assumption
of uniform distribution of nodes in the embedding space, which is rather
unrealistic in the context of social systems. Moreover, they tell little
about small-worldness and degree distributions.
We now turn to computer simulations to address these problems.

\section{Social Distance Attachment model}\label{sec:sda}

\begin{figure}[!t]
    \centering
    \includegraphics[width=0.45\textwidth]{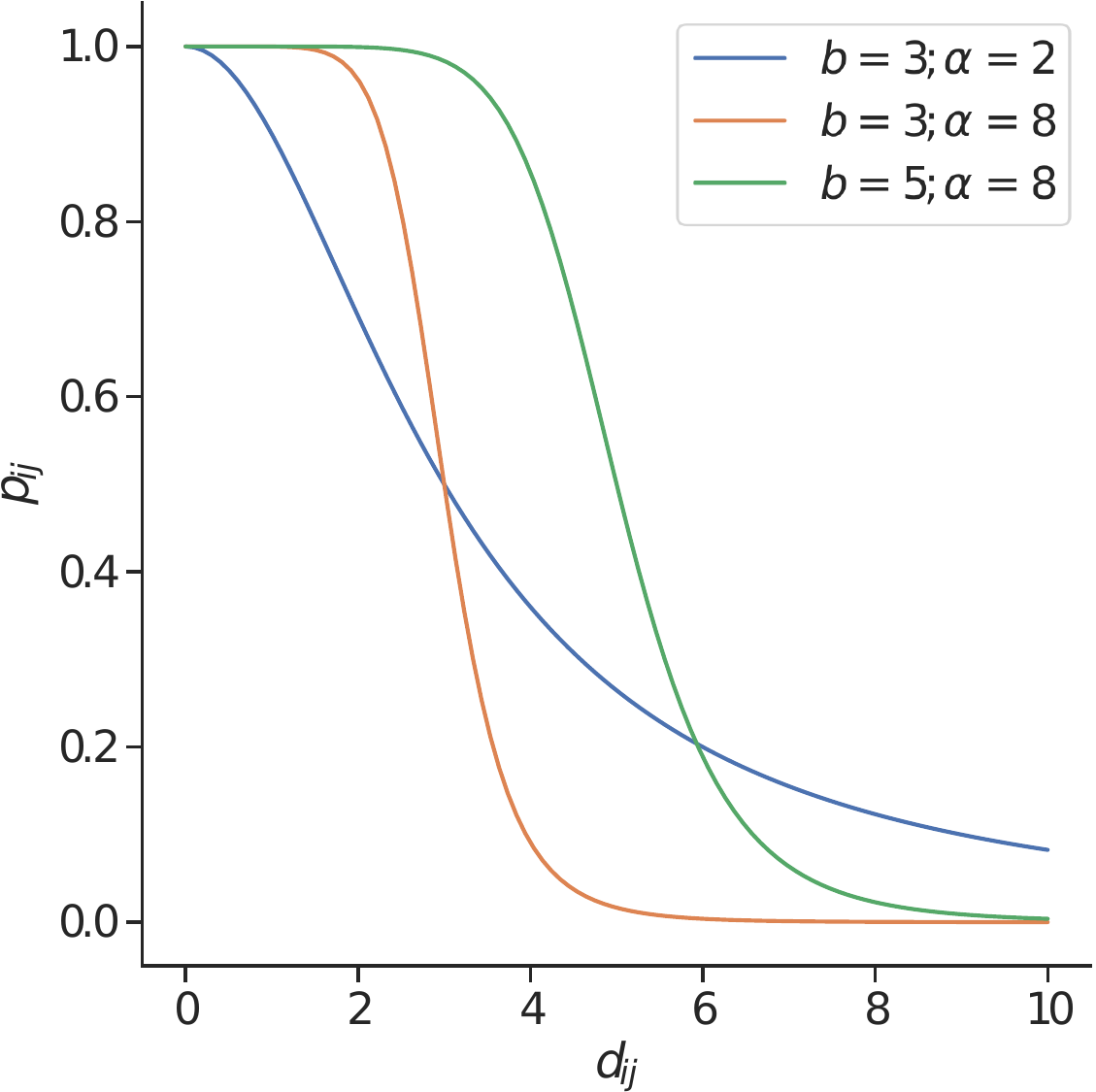}
    \caption{Connection probability function in SDA model.}
    \label{fig:sda}
\end{figure}

For the purpose of simulation studies we use the Social Distance Attachment
(SDA) model introduced by
\citet{boguna_models_2004}.
It is a RGG model formulated as follows.
Let $\Space_m = \R^m$ be an $m$-dimensional social space with an associated
distance metric $d_{ij} = d(\cdot, \cdot)$ (e.g. Euclidean or Manhattan distance)
and let $\mat{x}_i \in \Space_m$ for $i = 1, \ldots, N$ be points in this space.
Then, for all possible edges in an $N$-by-$N$ adjacency matrix (except self-loops)
the following connection probability is assigned:
\begin{equation}\label{eq:sda}
    p_{ij} = \frac{1}{1 + [b^{-1}d(\mat{x}_i,\mat{x}_j)]^\alpha}
\end{equation}
where $b$ is the characteristic distance and $\alpha$ is the level of homophily.
Similarity to our formulation of Fermi-Dirac distribution is not accidental,
since the SDA connection function is also a decreasing sigmoidal function
of distance in which $b$ and $\alpha$ play the same roles (see fig.~\ref{fig:sda}). \\

One remark is needed now. Why choose SDA function instead of
Fermi-Dirac distribution, which we used to show the connection
between social networks, homophily and random geometric graphs?
The answer is two-fold.
SDA function has a similar sigmoidal shape and also
converges to hard RGG model when $\alpha \to \infty$. Hence, being perhaps
slightly more biased since Fermi-Dirac distribution is the maximum entropy
solution, it is still qualitatively very similar. However, it also has two
very important practical advantages.

Firstly, assuming there are no zero distances,
for any $\alpha$ we have that $\E[k] \to 0$ as $b \to 0$ and
$\E[k] \to N-1$ as $b \to \infty$, while the former is not always true for
Fermi-Dirac distribution (see eq. \ref{eq:fd}). This makes it always possible
to find $b$ yielding any desired $\E[k]$ for any value of $\alpha$,
assuming that the distribution of nodes in an embedding space is not concentrated
in a limited number of discrete positions. Together with the fact
that conditionally on the embedding space edges in a network are independent
Bernuolli random variables and that the expected degree can be
easily computed as:
\begin{equation}\label{eq:Ek}
    \E[k] = \frac{1}{N}\sum_{i=1}^N\sum_{j=1}^N p_{ij}
\end{equation}
it means that the approximate value of $b$ can be always found with simple
numerical methods such as the bisection method
\citep[section~2.1]{burder_numerical_2010}
\footnote{
    In practice one would want to use a more efficient algorithm such as TOMS 748
    \citep{alefeld_algorithm_1995},
    which is exactly the method we used in our implementation of the model.
}. This is a very important property of the SDA connection function since it
means that SDA model can generate networks with any expected average node degree
and any level of homophily, which is necessary to allow for arbitrary
levels of sparsity.

Secondly, in the SDA model $\alpha$ scales normalized unitless distance
$d_{ij} / b$ (see eq. \ref{eq:sda}),
so values assigned to $\alpha$ always correspond to the same
level of homophily regardless of the distribution of nodes in the social space
and its measurement scale.
This makes it very easy to compare models run on different datasets.

Summing up, SDA model with fixed $\alpha$ and $\E[k]$ can be
computed according to the following procedure:

\begin{enumerate*}
    \item Let $\Space_m$ be an $m$-dimensional social space and
    $\mat{x}_i$ for $i = 1, \ldots, N$ be points in $\Space_m$.
    \item Derive $N$-by-$N$ distance matrix $\mat{D}_N$ between all pairs of
    points from $\Space_m$ using some distance metric.
    \item Choose values of $\alpha$ and $\E[k]$.
    \item Find $b$ using any univariate numerical root finding algorithm such
    as the bisection method. The objective is to find the root of the function:
    $$
    f(b) = \E[k] - \frac{1}{N}\sum_{i=1}^N\sum_{j=1}^N p_{ij}(b)
    $$
    Note that $p_{ij}$ (eq.~\ref{eq:sda}), given a distance between nodes $i$ and $j$,
    depends only on $b$ since we fixed $\alpha$. Thus, conditional on
    $\mat{D}_N$, $\alpha$ and $\E[k]$ this is a one-dimensional problem and
    thanks to the properties of the SDA connection probability function
    we can always solve it, because the root is unique and always exists.
    \item Transform $\mat{D}_N$ into a connection probability matrix
    $\mat{P}_N = (p_{ij})$. Note that $\mat{P}_N$ specifies a probability
    distribution over all networks with $N$ nodes.
    \item Use $\mat{P}_N$ to generate undirected or directed adjacency matrices.
    Every edge is created independently
    (since they are conditioned on the social space)
    with probability $p_{ij}$.
\end{enumerate*}

\section{Simulation study of Social Distance Attachment model}\label{sec:sda-sim}

In this section we study the behavior of SDA model with the Euclidean distance
metric in regard to clustering, degree assortativity, average path lengths
and degree distributions based on different underlying social spaces
(different probability distributions used to sample nodes' coordinates) and
a wide range of parameters' values. We also consider the behavior of the model
when combined with a random edge rewiring process in which each edge is destroyed
with some small probability and then a randomly selected node at one of its ends
creates a new link picking a new neighbor uniformly at random from
the pool of disconnected nodes. It is introduced to check whether such a small
random disturbance may be enough to ensure small-worldness while not affecting
the general effects of homophily with respect to clustering and degree
assortativity. We considered the following parameter values:
\begin{itemize*}
    \itemsep0em
    \item \textbf{system size ($N$):} $1000, 2000, 4000, 8000$
    \item \textbf{social space:}
    \vspace{-0.5em}
    \begin{itemize*}
        \itemsep0em
        \item uniform
        \item $4$ Gaussian clusters (equally sized spherical groups formed
        as multivariate Gaussian noise with different centroids)
        \item lognormal (coordinates sampled from independent lognormal distributions)
    \end{itemize*}
    \vspace{-.2em}
    \item \textbf{dimensionality of social space ($m$):} $1, 2, 4, 8, 16$
    \item \textbf{homophily ($\alpha$):} $2, 4, 8, \infty (\text{hard RGG})$
    \item \textbf{expected average degree ($\E[k]$):} $30$
    \item \textbf{probability of random rewiring (}$p_{\text{rewire}}$\textbf{):}
        $0, 0.01$
\end{itemize*}
Every combination of simulation parameters was run 10 times
(2 independent realizations of a social space and
5 independent realizations of an adjacency matrix).
Confidence intervals on plots show minima and maxima estimated based on
100 bootstrap replicates. Social spaces were generated with standard
pseudo-random numbers generators. This totaled to 4800 simulation runs.
Appendix~\nameref{app:sda} presents example realizations of social space and
corresponding networks.

The parameter space we explore is much richer than the one used by
\citet{boguna_models_2004}.
In particular, we study in detail distributions of nodes in social
spaces other than uniform as well as behavior of SDA in high dimensions,
while the original paper covers only the case of 1D uniform distribution.
Our selection of distributions is of course somewhat arbitrary, but it also
covers important qualitatively distinct cases.
Uniform distribution must be included
as a well-studied canonical benchmark, Gaussian clusters correspond to one of
the most typical geometries of a structure with clear grouping
(spherical clusters around different centroids) and lognormal is a very
common type of a right-skewed distribution with a fixed lower bound. It is
often used to model such quantities as income and arises naturally as a limiting
distribution of multiplicative quantities, since it is a result of applying
Central Limit Theorem to the logarithm of the product of a sequence of values.

All graph-theoretic quantities were computed with \textit{igraph}
\citep{csardi_igraph_2006}.
R language
\citep{r_core_team_r:_2019}
was used for data analysis and visualizations.

\begin{figure}[!t]
    \centering
    \includegraphics[width=.8\textwidth]{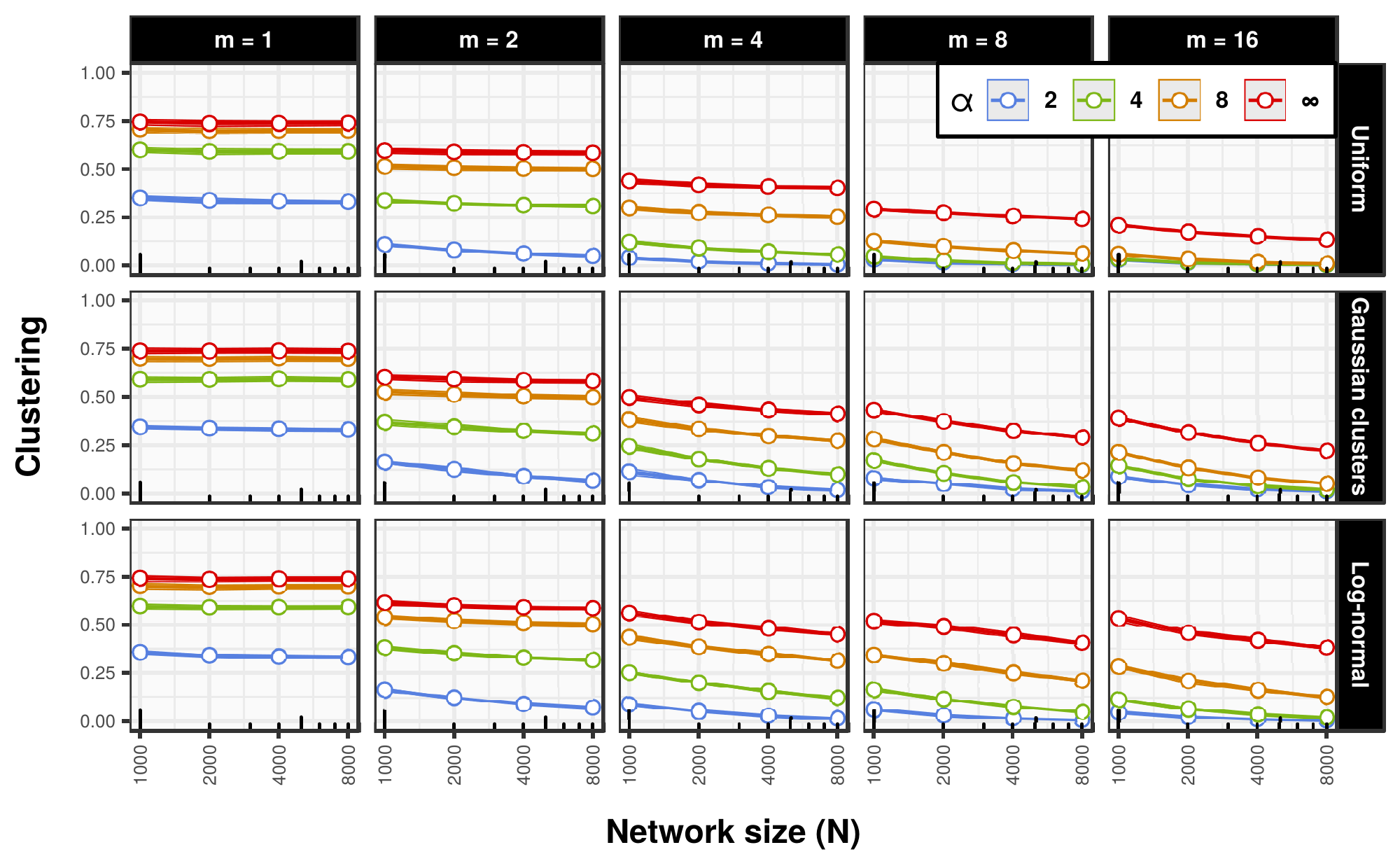}
    \caption{
        SDA simulation results for clustering.
        The results are averaged over networks with and without the random rewiring.
        Confidence bounds show minima and maxima estimated based on 100
        bootstrap replicates.
    }
    \label{fig:sda-cl}
\end{figure}

The simulations showed that for all of the embedding spaces clustering increases
with $\alpha$ and slightly decreases with system size and corresponding
lower density of connections (see fig.~\ref{fig:sda-cl}). The relationship is
approximately linear with respect to $\log N$, but in low dimensional spaces
($m = 1$ and $m = 2$) the level of clustering is approximately constant for
all system sizes. Furthermore, clustering decreases markedly
in higher dimensions and this effect seems to be strongest when nodes are
distributed uniformly.
This is consistent with the existing analytical results for RGGs with uniformly
distributed nodes that show that
clustering, while non-vanishing in the limit of large system size, decreases
towards zero with the increasing number of dimensions of the embedding space
\citep{dall_random_2002}.
It suggests that homophily-induced clustering is more
robust in higher dimensions when there is a natural grouping of nodes in the
embedding space. Results are averaged jointly over networks with and without
the random rewiring, but in all cases confidence bounds are
very narrow indicating that there is little variation between different
network realizations.

\begin{figure}[!t]
    \centering
    \includegraphics[width=.8\textwidth]{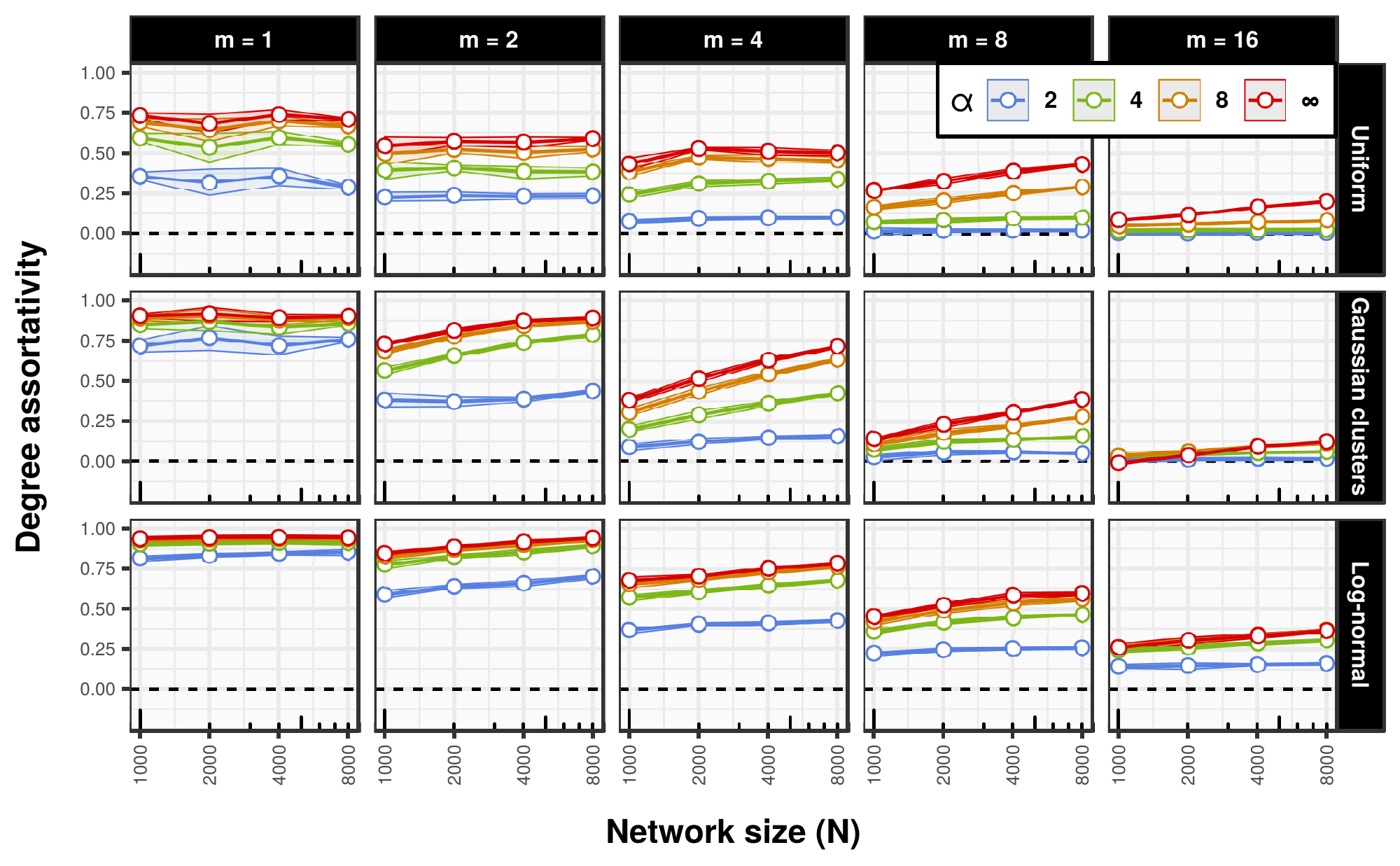}
    \caption{
        SDA simulation results for degree assortativity.
        The results are averaged over networks with and without the random rewiring.
        Confidence bounds show minima and maxima estimated based on 100
        bootstrap replicates.
    }
    \label{fig:sda-as}
\end{figure}

Assortativity (see fig.~\ref{fig:sda-as}) also increases with $\alpha$, but it
also increases with $N$, especially in higher dimensional spaces. The relationship
is approximately linear with respect to $\log N$.
On average it also decreases with the dimensionality of an embedding space.
Again, narrow confidence bounds show little variability between different
network realizations and no significant differences between networks with
and without the random rewiring.

\begin{figure}[!t]
    \centering
    \includegraphics[width=.8\textwidth]{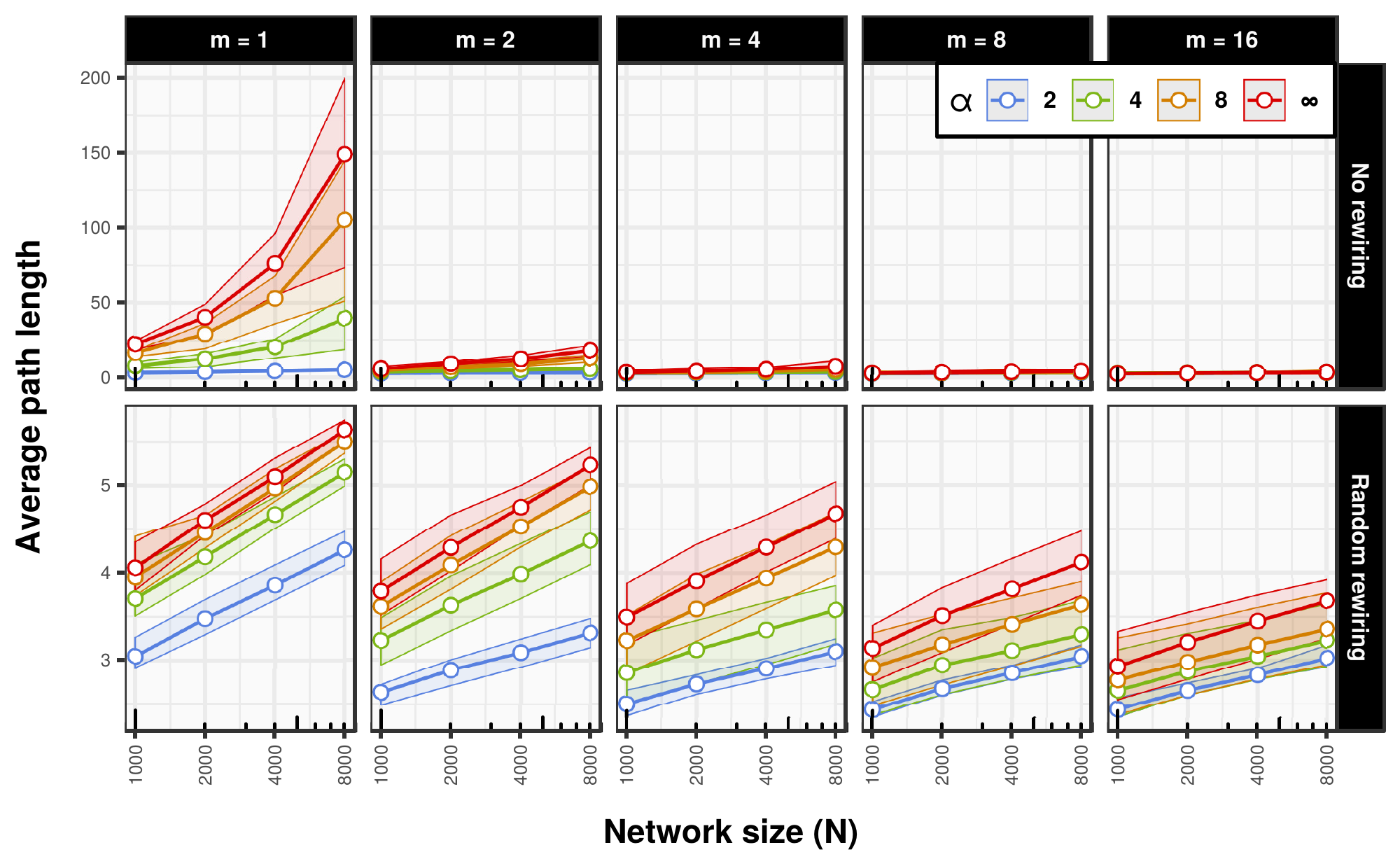}
    \caption{
        SDA simulation results for average path lengths.
        Confidence bounds show maxima and minima estimated based on 100
        bootstrap replicates.
    }
    \label{fig:sda-L}
\end{figure}

Average path lengths grow superlinearly with $\log N$ in some cases when
there is no random rewiring (see fig.~\ref{fig:sda-L}).
When the random rewiring is present average path lengths
grow linearly with $\log N$ in all cases. This shows that homophily by itself,
especially when it is strong, does not yield the small-world effect.
It suggests that in real social systems homophily-induced assortative mixing
may be accompanied by more random-like processes creating some small number
of relationships that go beyond structural constraints based on the structure of
the social space. This connects very well to the seminal results concerning
the importance of weak ties for connectivity and access to as well as
distribution of resources in social networks
\citep{granovetter_strength_1973}.

To assess the extent to which networks with the random rewiring conform to the
small-world scaling we computed Pearson correlations between average path lengths
and $\log N$ for all combinations of parameters. Minimum correlation was
0.969, maximum 1 and median was 0.996. The scaling is almost perfectly linear,
so we conclude that random rewiring ensures the small-world property in
homophily-driven networks in all cases. Moreover, very narrow confidence bounds
for clustering and assortativity show that the random edge rewiring does not affect
these properties in any meaningful way, so it can be mixed with homophily
without distorting or attenuating its main effects.

\begin{figure}[!t]
    \centering
    \includegraphics[width=.8\textwidth]{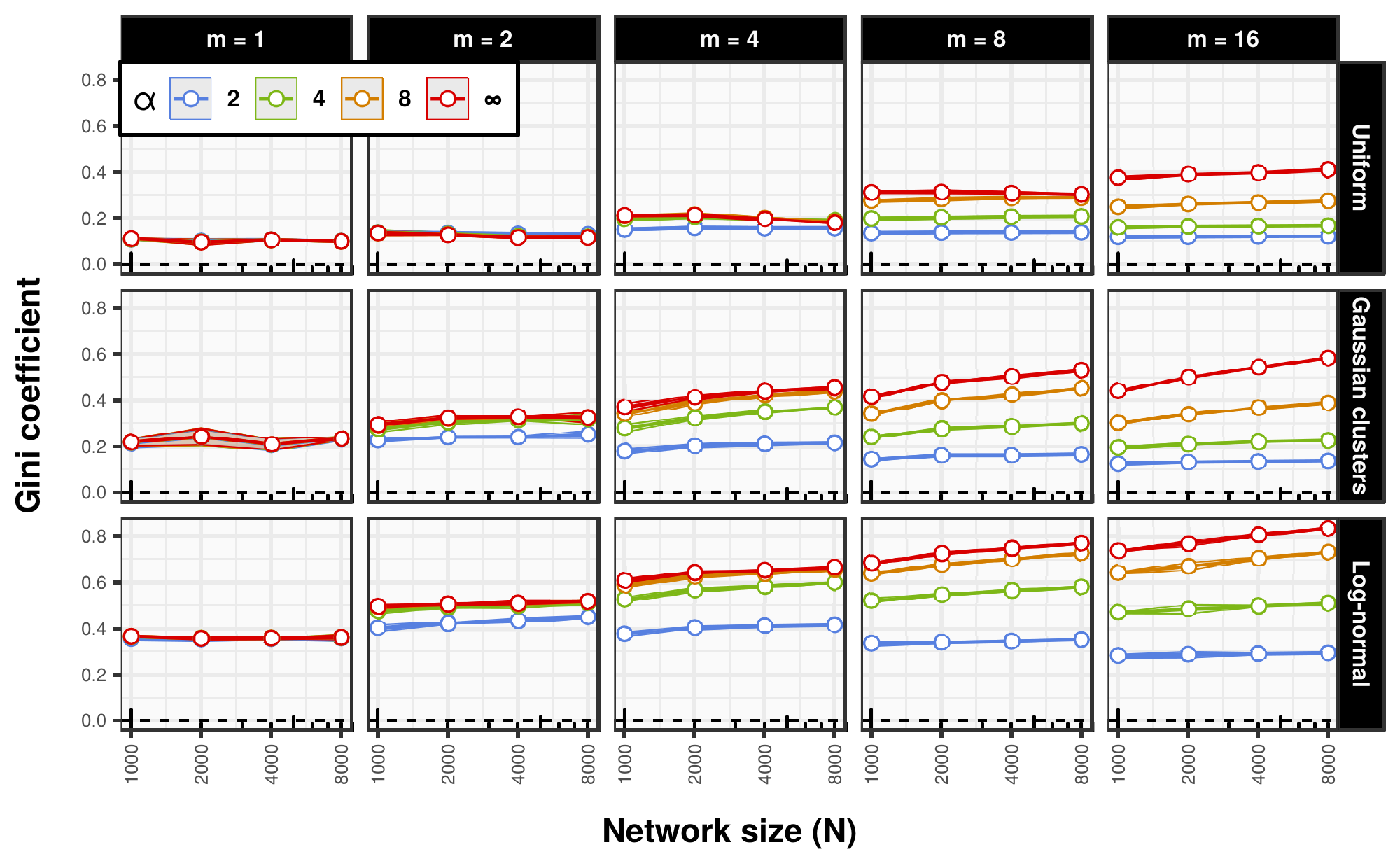}
    \includegraphics[width=.8\textwidth]{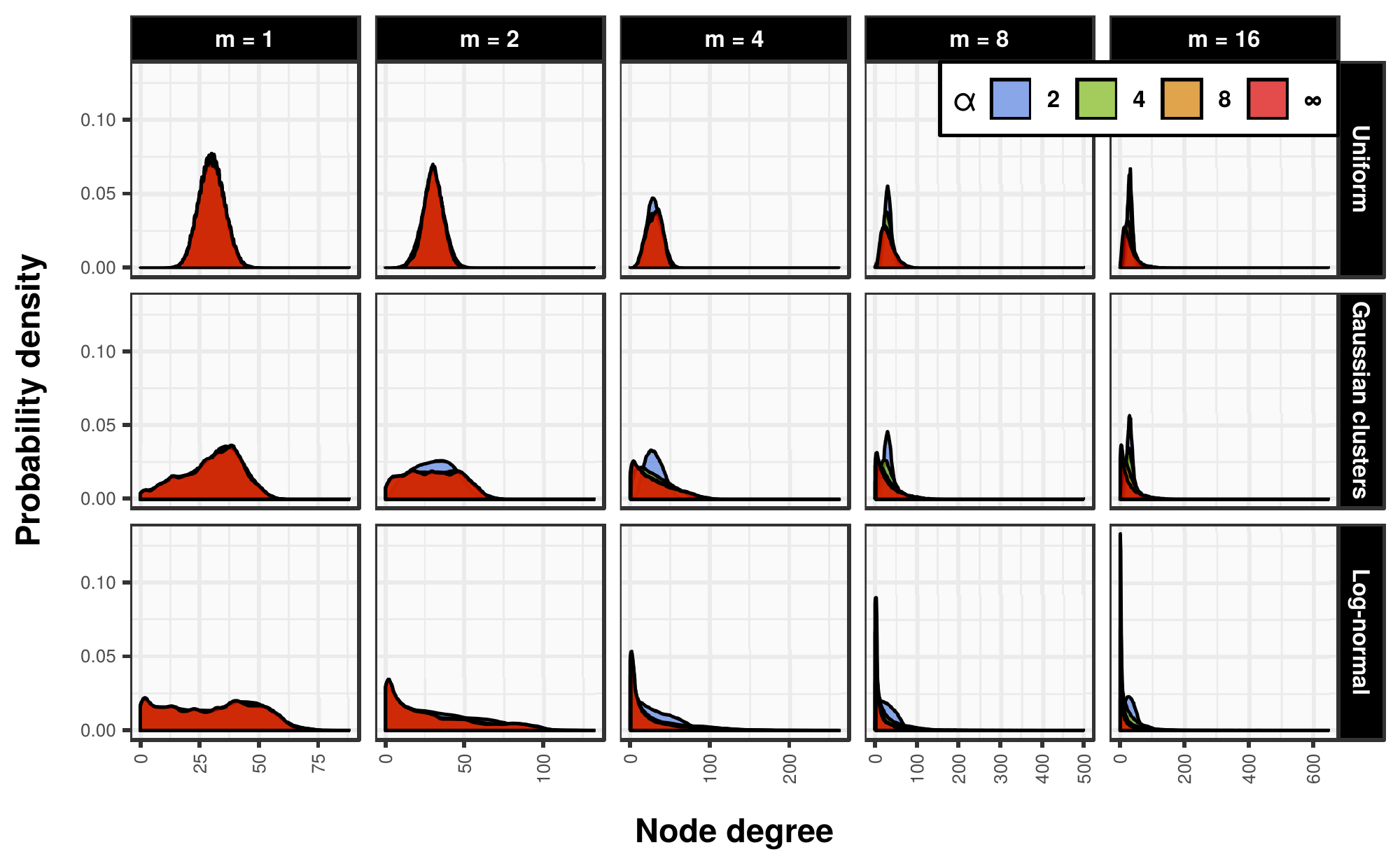}
    \caption{
        (top) SDA simulation results for Gini coefficient of degree distributions.
        Confidence bounds show maxima and minima
        estimated based on 100 bootstrap replicates.
        (bottom) Averaged degree distributions for $N = 8000$ simulated in SDA model.
    }
    \label{fig:sda-dists}
\end{figure}

Diversity of degree distributions in terms of inequality (right-skewness)
as measured with Gini coefficient
\citep{badham_measuring_2013}
clearly depends on the structure of social space (see~fig.~\ref{fig:sda-dists}).
The more clustered and concentrated the space is the more unequal is the
distribution, so it is lowest for uniform social spaces and highest for
lognormal ones. Also dimensionality of a space increases values of Gini
coefficient. On the other hand the strength of homophily seems to matter only
in higher dimensional spaces where it increases inequality when it is high and
decreases when it is low. Also there is a clear linear scaling of inequality with
respect to $\log N$ in higher dimensions when homophily is strong. In general
the results suggest that homophily is sufficient to generate a~quite diverse
range of degree distribution shapes.

Now we turn to the problem of boundedness of degree distributions. We have
seen that homophily may lead to right-skewed degree distributions, but it is
still unclear to what extent it may produce scale-free distributions.
The Dunbar's numbers hypothesis implies that strong power law
degree distributions should be rare in social networks, so it is worthwhile
to assess the prevalence of scale-free distributions in networks generated
with SDA model. To do so, we use a statistical approach based on
Extreme Value Theory developed by
\citet{voitalov_scale-free_2019}.
Our analysis showed that the prevalence of scale-free distributions in the
simulated networks is really low. 4732 networks (98.6\%) were classified as
'not power laws' (no statistical evidence of a power law tail) and 68 (1.4\%)
were classified as 'hardly power laws'
(very little evidence of a power law tail) based on the classification
scheme developed by the authors of the methodology. Hence, we conclude
that homophily is unlikely to be sufficient to generate power law
degree distributions.

\subsection{Summary}\label{sec:sda-summary}

In this section we presented and discussed an extensive simulation study of
SDA model with respect to a variety of parameters' configurations and embedding
social spaces. The results confirm that SDA can produce networks
with high clustering and positive degree assortativity that do not vanish
in the limit of large system size. Therefore homophily may play an important
role in the processes generating social networks for it reproduces the crucial
properties of clustering and assortativity and at the same time has a natural
sociological interpretation as well as has been frequently observed in
empirical data.
We also showed that (hard) RGG model properly describes the limiting behavior
of SDA model when $\alpha \to \infty$ and as such it can be used to deduce
an approximate behavior of SDA with strong homophily.

Moreover, we showed that homophily alone does not guarantee the small-world
property. In fact they can be in a direct conflict,
especially when homophily is strong
and the dimensionality of an embedding space is low.
However, technically this problem can be remedied quite easily by
randomly rewiring a small fraction of edges.
Yet more importantly, it indicates that homophily,
if occurs, is likely to be mixed with other random processes allowing some
edges to go beyond structural constraints imposed by it.

Finally, right-skewness (inequality) of degree distributions clearly increases
with system size and especially dimensionality of an embedding social space,
so in general homophily is sufficient to generate a diverse range
of degree distributions. Nevertheless, we found no evidence of scale-free
distributions in the generated degree sequences, but applicability of this
result is limited only to the boundaries of the parameters' space that was explored.
Moreover, the results indicate that degree distributions are to a large extent
determined by geometry of an embedding social space. Together with the fact
that levels of clustering and degree assortativity are relatively robust
with respect to nodes' position, it implies that the control
over the distribution of positions in a~social space allows some level of
an indirect control over the shape of the degree distribution.

\section{Social Distance Configuration model}\label{sec:sdc}

The configuration model which allows generating networks with arbitrary degree
distributions, at least approximately if one corrects for multiple edges and
self-loops, can be extended to produce also prespecified numbers of triangles
and as a result arbitrary levels of clustering
\citep{newman_random_2009}.
However, its formulation is strictly technical and algorithmic and as such
can hardly be interpreted in sociological terms. Hence, it can be very useful
analytically, but it is unclear what kind of theoretically significant social
process it could represent. To~address this issue we propose a hybrid model
combining the standard configuration model with SDA model.
For now we call it the Social Distance Configuration (SDC) model.

The idea is very simple. The algorithm of the  ordinary configuration model can
be defined as follows
\citep[p.~435]{newman_networks:_2010}:
\begin{enumerate*}
    \item Take a degree sequence $k_1, k_2, \ldots, k_n$
    (it must sum to an even number) and assign a degree $k_i$ to each node.
    At this point node degrees represent not edges but so-called stubs
    or half-links.
    \item Choose a node uniformly at random from the set of all nodes with
    non-zero number of stubs and decrease its stub-degree by one.
    \item Choose another node in the same fashion and decrement its stub-degree.
    Note that it may be the same node.
    \item Connect the two selected nodes with an edge. Note that any
    given pair may be connected by multiple edges.
    \item Repeat steps 2-4 until there are no more free stubs.
\end{enumerate*}

The trick that we propose is to first compute edge creation probabilities
according to SDA model and then sample pairs of nodes not uniformly at random
but with probabilities proportional to corresponding edge creation probabilities.
This achieves the objective of the configuration model, but does so in a way that
privileges edges between nodes that are close to each other in an embedding
social space. As a side effect, it allows setting very low connection
probabilities (but necessarily non-zero) to multiple edges and self-loops so
they can occur as rarely as possible. The pseudo-code is presented in the
appendix~\nameref{app:sdc-algorithm}.

The sociological rationale for this approach is the following. Connection
probabilities are derived from SDA, which itself is based on the clearly
interpretable notion of homophily. On this process we enforce an arbitrary
degree sequence, of which interpretation depends on the context. For instance,
preferential attachment can be interpreted in various social ways such as
rich-get-richer mechanisms. At the same time it is associated with a well-defined
degree distribution --- a power law with the exponent $\gamma \approx 3$.
Hence, SDC may be used, for instance, to simulate
preferential attachment process embedded in a social space with homophily.
In general, SDC model makes it possible to study effects of homophily
(as operationalized in SDA model) while allowing degree distributions
to be constrained by some other processes.

In the next section we present the results of a simulation study of the
behavior of SDC with respect to a variety of parameters' values, social spaces
and degree sequences. This will enable us to further explore properties of
homophily as a social network generating process and examine its
robustness under strict constraints imposed on degree distributions.

\section{Simulation study of Social Distance Configuration model}\label{sec:sdcm-sim}

The setting is analogous to the simulation study of SDA model. We examine
the behavior of SDC with the Euclidean distance metric with respect to clustering,
degree assortativity and average path lengths under three enforced degree
distributions: Poisson, negative binomial ($n = 1$; $p = 1 / 31$)
and discrete power law (characteristic exponent $2 \leq \gamma \leq 3$).
All degree sequences
are generated in such a way that their expected average values equal 30, which
is the average degree simulated previously. Simulating Poisson and negative
binomial sequences with fixed $\E[k]$ is a trivial task, but power laws
pose some technical difficulties. We describe our approach to simulating
power law graph degree sequences with fixed $\E[k]$
in the appendix~\nameref{app:simulate-pl}. The types of degree distributions
were selected to cover three very typical qualitatively different cases:
\begin{enumerate*}
    \itemsep0em
    \item Approximately symmetric distribution with finite moments (Poisson).
    \item Right-skewed distribution with finite moments (negative binomial).
    \item Right-skewed distribution with infinite moments (power law).
    Even though the prevalence of power law degree distributions in social networks
    is debatable, this case is still important for the study as it provides
    an ideal type for a strongly right-skewed distribution (dominated by few hubs),
    which is qualitatively different from more ,,well-behaved'' right-skewed
    distributions with finite moments.
\end{enumerate*}

The same space of parameters' values was explored, only this time we did not
explore differences between networks with and without rewiring
($p_{\text{rewire}}$ is set to 0.01 for all cases). All together the following
parameters were used (see appendix~\nameref{app:sdc} for network examples):
\begin{itemize*}
    \itemsep0em
    \item \textbf{degree sequence:} Poisson, negative binomial, power law
    \item \textbf{system size ($N$):} $1000, 2000, 4000, 8000$
    \vspace{-0.5em}
    \begin{itemize*}
        \itemsep0em
        \item uniform
        \item $4$ Gaussian clusters (spherical groups formed
        as multivariate Gaussian noise with different centroids)
        \item lognormal (coordinates sampled from independent lognormal distributions)
    \end{itemize*}
    \vspace{-.2em}
    \item \textbf{dimensionality of social space ($m$):} $1, 2, 4, 8, 16$
    \item \textbf{homophily ($\alpha$):} $2, 4, 8, \infty (\text{hard RGG})$
    \item \textbf{expected average degree ($\E[k]$):} $30$
    \item \textbf{probability of random rewiring (}$p_{\text{rewire}}$\textbf{):}
        $0.01$
\end{itemize*}
Every parameters' configuration was run 12 times
(2 independent realizations of a social space and 6 independent realizations
of a degree sequence). This totaled to 8639 realizations
(one run was dropped due to computational issues).

The general qualitative behavior of SDC with respect to clustering is similar
to SDA (see~fig.~\ref{fig:sda-cl}). In low dimensional spaces it is approximately
constant regardless of system size, while in higher dimensional spaces it
decreases approximately linearly with $\log N$. However, general levels of
clustering are markedly lower than in the case of SDA and hard RGG
($\alpha = \infty$) often exhibit lower clustering than SDC realizations with
finite $\alpha$. This is due to the fact that in hard RGG all nodes beyond the
characteristic distance $b$ look the same from the vantage point of connection
probability (see appendix~\nameref{app:sdc-algorithm}).
This makes it more likely that some
of the edges will be purely random and as result decreases clustering
and assortativity. This is a computational artifact with no meaningful
social interpretation, so perhaps it is preferable to run SDC with finite
values of $\alpha$.

\begin{figure}[!t]
    \centering
    \includegraphics[width=.8\textwidth]{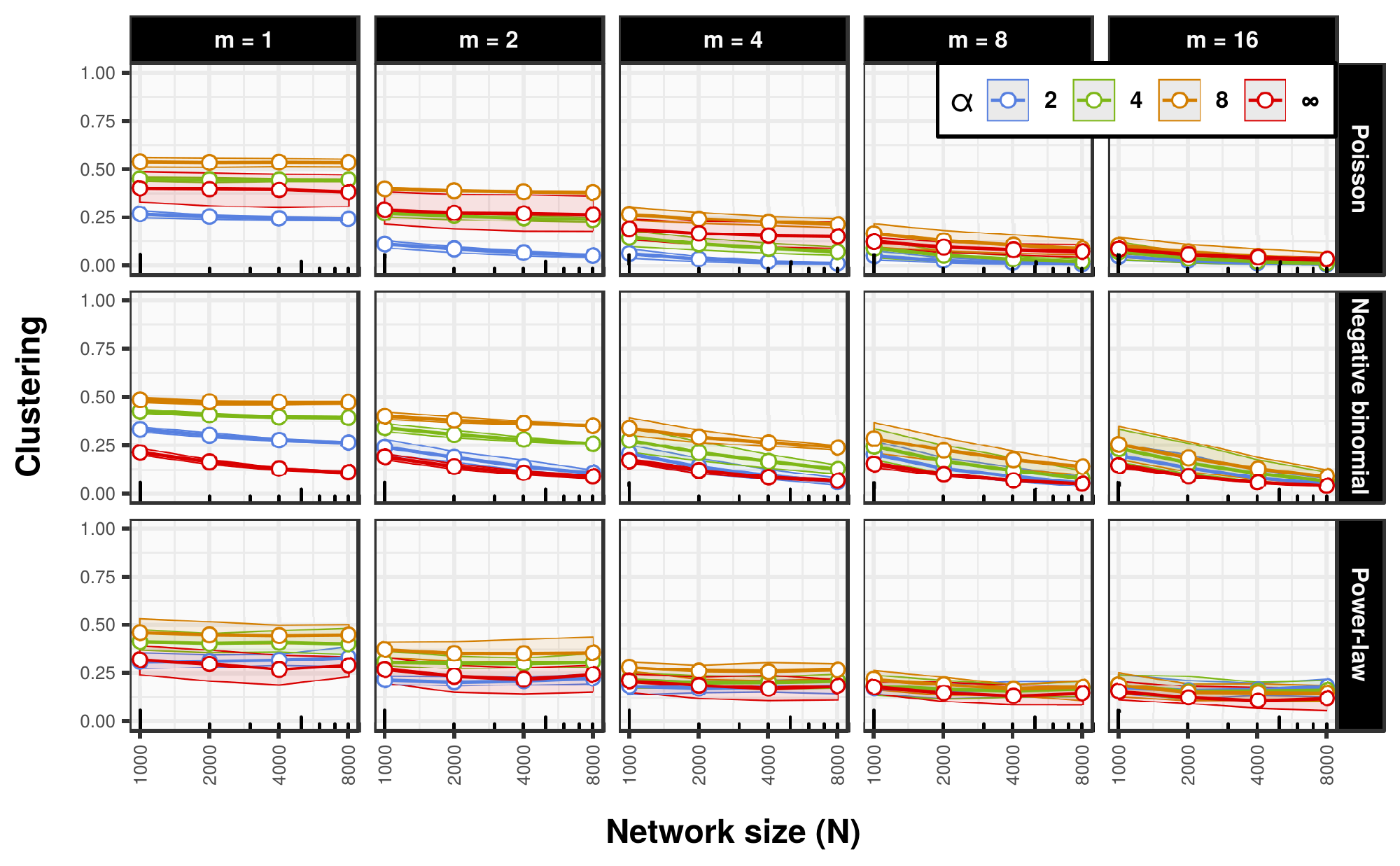}
    \caption{
        SDC simulation results for clustering.
        Confidence bounds show minima and maxima estimated based on
        100 bootstrap replicates.
    }
    \label{fig:sdc-cl}
\end{figure}

\begin{figure}[!t]
    \centering
    \includegraphics[width=.8\textwidth]{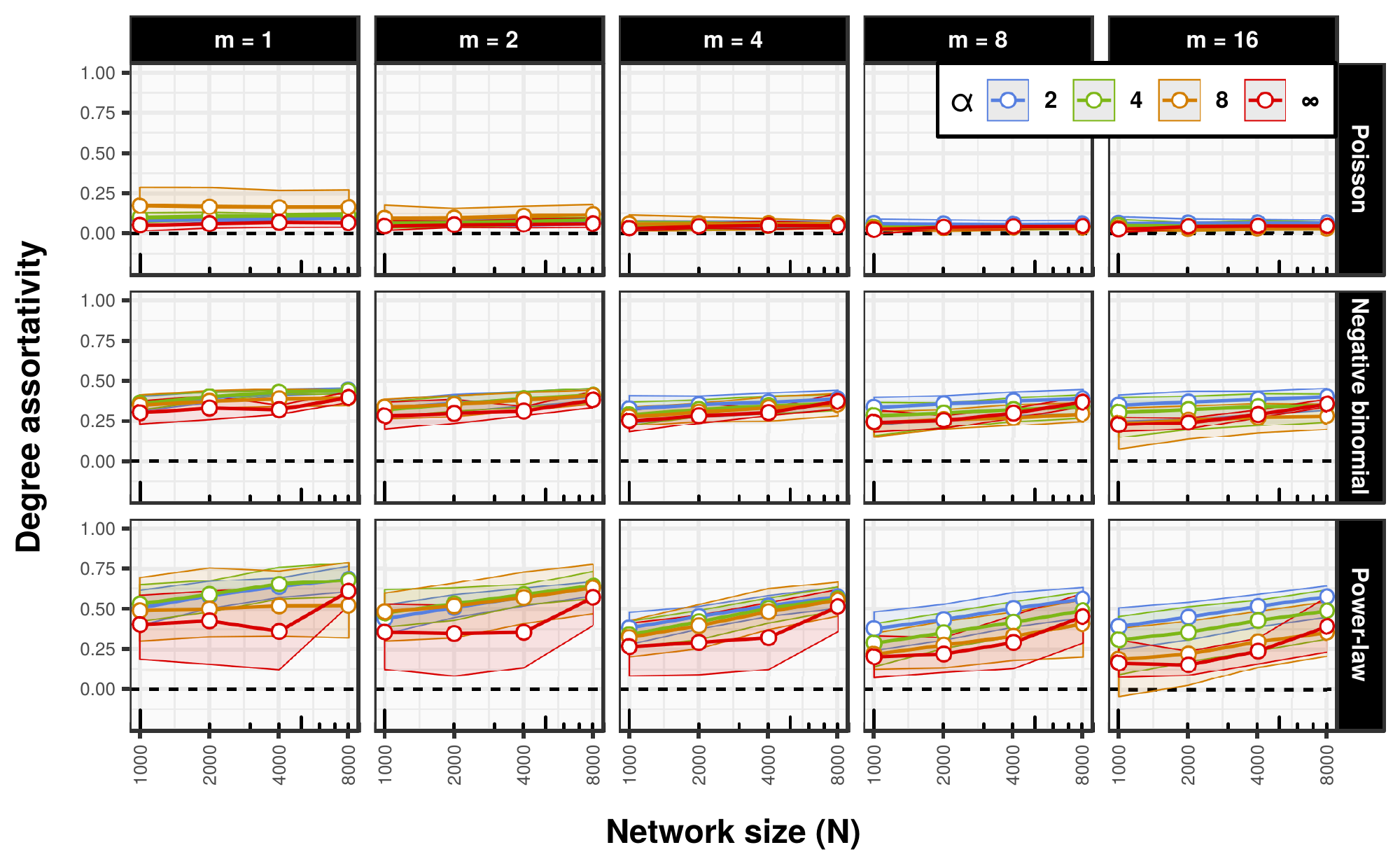}
    \caption{
        SDC simulation results for degree assortativity.
        Confidence bounds show minima and maxima estimated based on
        100 bootstrap replicates.
    }
    \label{fig:sdc-as}
\end{figure}

With respect to degree assortativity the behavior of SDC is more complicated
and there are some important differences between types of degree sequences
(see~fig.~\ref{fig:sdc-as}).
In particular, it should be noted that in almost all cases assortativity for
Poisson degree sequences is near zero and remains stable irrespective
of the size of a system or the dimensionality of a social space,
while in the case of negative binomial distributions it is approximately fixed
around the value of $0.3$. On the other hand power law sequences have
more variability, but in general effects of the strength of homophily ($\alpha$)
are quite weak. This, combined with the SDA model results concerning
assortativity, suggests that the influence of $\alpha$ on degree assortativity
is largely dependent on the ability of homophily to reorganize degree distributions,
so when they are fixed the influence decreases. On the other hand, also when
combined with degree distribution constraints, homophily still seems to be unable
to generate networks with negative assortativity, which points to
the robustness of its connection to non-negative degree correlations.

\begin{figure}[!t]
    \centering
    \includegraphics[width=.8\textwidth]{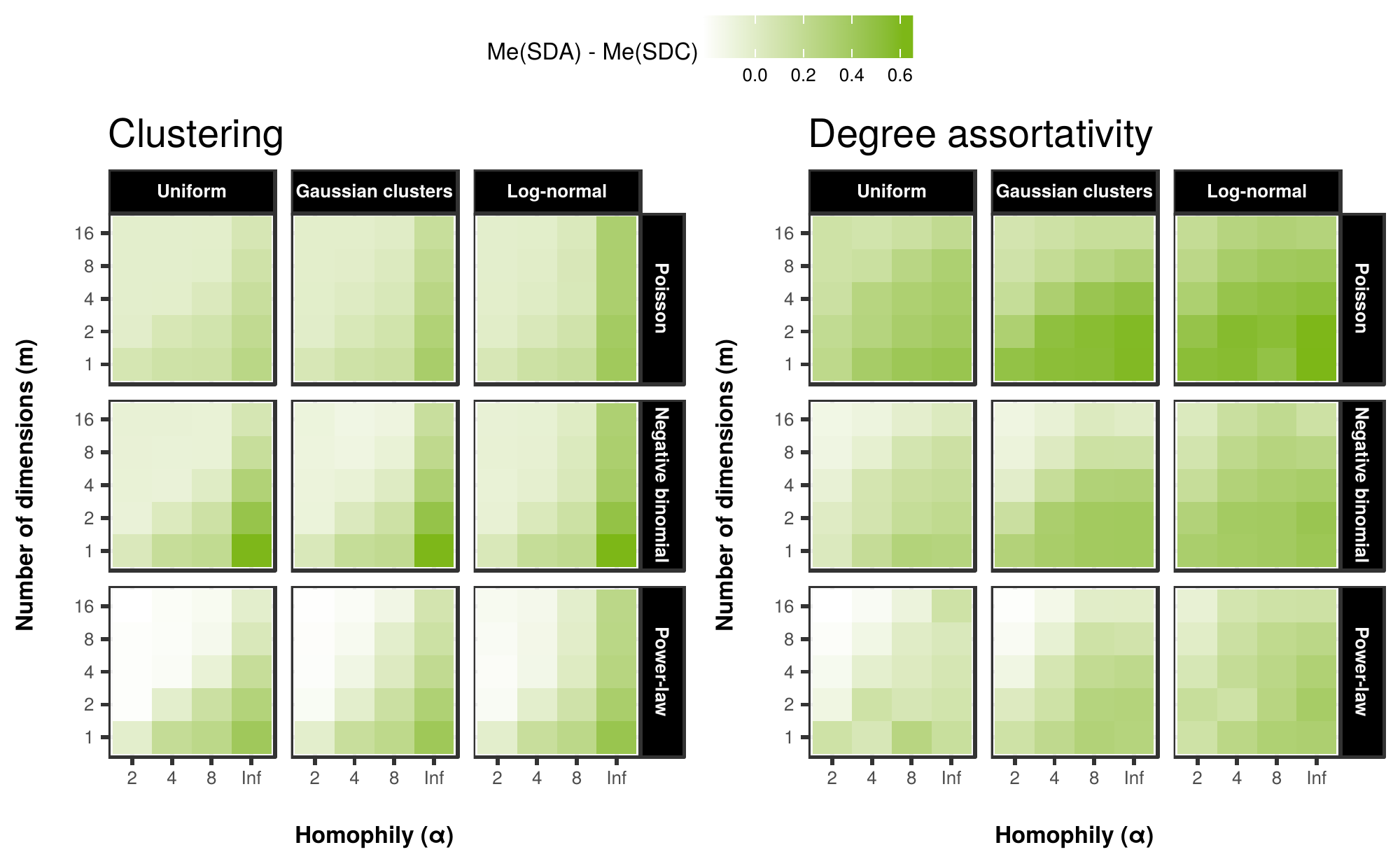}
    \caption{
        Average differences between median levels of clustering (left)
        and assortativity (right) in SDA and SDC models. Medians were computed
        for each combination of parameters' values
        (plots show aggregated values broken down by social space, degree sequence,
        dimensionality and homophily).
        White tiles denote differences that were on average insignificant.
        Significance was assessed with Wilcoxon Rank Sum test (p-values were
        corrected for multiple testing with Benjamini-Hochberg-Yekutieli FDR method).
    }
    \label{fig:sda-sdc-comparison}
\end{figure}

Additionally, we assessed differences between median values of clustering
and assortativity in SDA and SDC networks
(see fig.~\ref{fig:sda-sdc-comparison} for details).
We found that in most of the cases SDC networks have both lower
clustering and degree assortativity. The differences are largest for
Poisson degree sequences and smallest for power laws. They also tend
to be more pronounced when homophily is strong and slightly weaker
in higher dimensional social spaces.
This confirms that constraints on degree distributions attenuate
the effects of homophily, although they do not cancel them completely.
Thus, the results show that homophily as a network generating process may
operate alongside other processes that constrain degree distributions
without loosing its crucial influence on clustering.

\begin{figure}[!t]
    \centering
    \includegraphics[width=.8\textwidth]{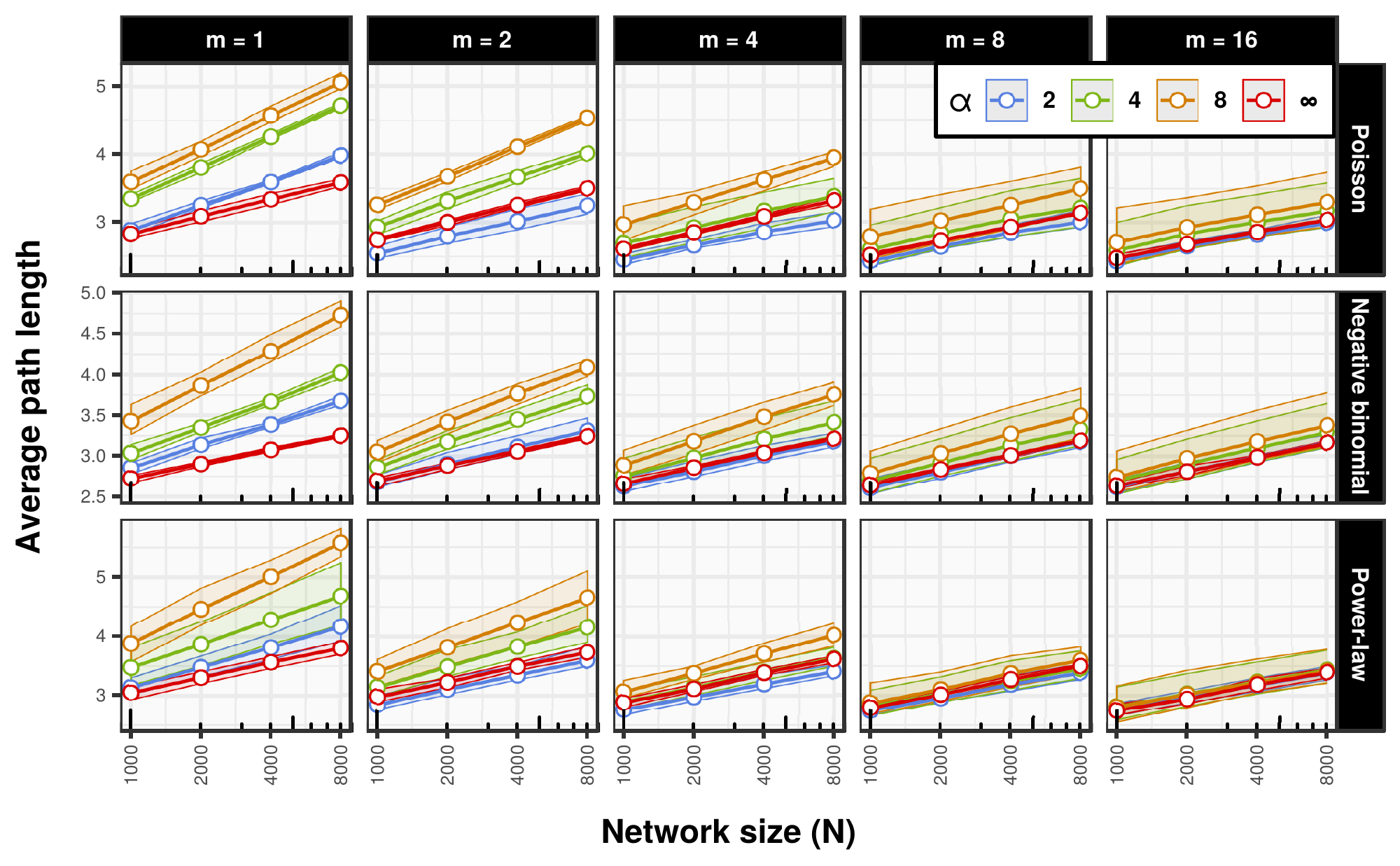}
    \caption{
        SDC simulation results for average path lengths.
        Confidence bounds show minima and maxima estimated based on
        100 bootstrap replicates.
    }
    \label{fig:sdc-L}
\end{figure}

Figure~\ref{fig:sdc-L} shows the relationship between average path lengths
and $\log N$ in SDC model with the random edge rewiring. We see that in all cases
the scaling is almost perfectly linear
(bootstraped $r \in [0.976, 1]$ with median of 0.999).
Therefore, the process of random edge
rewiring provides the small-world effect. Again, we see surprising results
in the case of $\alpha = \infty$, which tend to have lowest average path lengths.
This is caused by the same computational issue as the anomalies concerning
clustering and assortativity.

\subsection{Summary}\label{sec:sdcm-sim-summary}

In this section we introduced a hybrid model combining SDA and the configuration
model that allows generating social distance attachment networks with
arbitrary degree distributions. We showed that in general the model behaves
similarly to SDA, but in some cases its exact behavior strongly depends
an imposed degree sequence. The results have also more theoretical implications.
They show that homophily may operate in parallel to other process that impose
constraints on degree distributions and still yield significantly positive
clustering and in some cases positive assortativity as well,
although this property seems to depend on degree distribution to a much larger
extent. Therefore, homophily-induced clustering appears to be relatively robust
with respect to degree distribution constraints.
This provides further evidence that homophily may be an important process
shaping real-world social networks, also when interwoven with other processes.

\section{Discussion}\label{sec:discussion}

Our results based on the Social Distance Attachment (SDA) model show that
homophily can generate sparse networks with non-trivial clustering,
positive degree assortativity and diverse degree distributions, including
approximately symmetric as well as markedly right-skewed ones but no power laws.
Hence, the principle of homophily is a sufficient condition for most of the
typical properties of social networks to emerge. The results are consistent
with a rich body of research indicating that homophily links
social structure and social networks and shapes network structures
\citep{blau_macrosociological_1977,marsden_core_1987,popielarz_edge_1995,mcpherson_birds_2001,kossinets_origins_2009,centola_choosing_2015,halberstam_homophily_2016}.
In our analysis we considered what we call the static view of homophily
(see~section~\ref{sec:geometry}).
In general, similarity between connected agents
may be both due to selection (similarity breeds connection) or to social influence
(connection breeds similarity)
\citep{anagnostopoulos_influence_2008,flache_models_2017}.
However, the principle of homophily in its most general form is a statement
about a correlation which just implies that on average connected agents are more
similar than disconnected ones. Therefore we argued that it is important to study
typical network structures that such a~probabilistic pattern leads to,
since if homophily is really an important force shaping networks,
then such structures should be also relatively common in observed social
networks. We showed that networks generated by homophily
have most of the typical structural features of many social networks,
so our results complement existing empirical evidence of the link between
homophily and the structure of social networks.

Our analysis was based on formal models and computer simulations. This allowed
us to study effects of both pure homophily and homophily mixed with other
processes in a controlled and isolated manner. Through this we were able to
show that the small-world property
\citep{watts_collective_1998}
is not implied by homophily. In fact they can be in a direct conflict, especially
when the strength of homophily is high and/or the social space is simple,
that is, has few dimensions. However, we also showed that if some small fraction
of edges is random then the small-worldness is guaranteed without destroying
the structure induced by homophily
(i.e. sparsity, high clustering, positive degree assortativity and degree distributions).
This suggests that other processes ensuring some degree of randomness of
social ties are necessary, which is consistent with the seminal results
concerning the importance of weak ties
\citep{granovetter_strength_1973}.

Our formal approach allowed us also to extend SDA model and combine it with
the configuration model
\citep{newman_random_2001}
to study effects of homophily with arbitrary degree sequence constraints.
Based on this we showed that clustering induced by homophily is relatively robust
and independent of a particular degree distribution. On the other hand
degree assortativity is strongly influnenced by the degree distributions, so
once a~distribution is fixed, assortativity is largely independent of
the strength of homophily.

Last but not least, we derived our formal model from first principles based
on the direct correspondence between the principle of homophily and
random geometric graphs
\citep{dall_random_2002}.
This makes our synthetic results much less arbitrary and means that they
can be considered an important addition to the large body of empirical results
concerning connections between social structure, homophily and social networks.
Let us also note that much of the theory of social structure
\citep{blau_macrosociological_1977}
is based on the assumption that the principle of homophily holds.
Our results showing that homophily is sufficient to generate most of the typical
structural properties of social networks provide an additional indirect validation
of the correctness of this assumption. It is important since most of the
empirical evidence for the principle of homophily stays at the micro level
of the observed similarities between related social agents
\citep[e.g.][]{marsden_core_1987,aiello_friendship_2012}.
Our results add to that a formal support at the macro level as they show
that the principle of homophily is also congruent with typical macrostructures
of social networks.

We also note that our results suggest that homophily is unlikely to
lead to significantly negative degree assortativity,
at least under circumstances similar to those we studied.
On the other hand of course not all social networks have positive degree
correlations
(see \cite{estrada_combinatorial_2011} for some examples).
Hence, it is possible that there is a broad class of networks for which
homophily can not be the main generating process. This allows us to formulate
a crude rule of thumb that may be helpful for assessing
homophily/lack of homophily in real-world social networks:
\begin{itemize*}
    \itemsep0.5em
    \item
    \begin{minipage}{.4\textwidth}
        \textbf{Clustering and positive assortativity}
    \end{minipage}
    \begin{minipage}{.1\textwidth}
        $\Longrightarrow$
    \end{minipage}
    \begin{minipage}{.4\textwidth}
        \textbf{Homophily can be \\ the leading generating process}
    \end{minipage}
    \item
    \begin{minipage}{.4\textwidth}
        \textbf{Near zero clustering and/or assortativity}
    \end{minipage}
    \begin{minipage}{.1\textwidth}
        $\Longrightarrow$
    \end{minipage}
    \begin{minipage}{.4\textwidth}
        \textbf{Inconclusive}
    \end{minipage}
    \item
    \begin{minipage}{.4\textwidth}
        \textbf{Significantly negative assortativity}
    \end{minipage}
    \begin{minipage}{.1\textwidth}
        $\Longrightarrow$
    \end{minipage}
    \begin{minipage}{.4\textwidth}
        \textbf{Homophily can not be \\ the leading generating process}
    \end{minipage}
\end{itemize*}
The 'rule' we propose is of course greatly oversimplified, but we believe it may
be useful as a first guess when trying to assess possible generating mechanisms
of a social network.

Even though we derived our model and formalization of the homophily principle
in a systematic and analytic manner, it can not be ruled out that some other
models would be more appropriate and yield different results and conclusions.
Therfore it would be worthwhile to check in future studies the robustness
of our results based on other possible formalizations of homophily-like processes.
However, some similar work has already been done and it points to the robustness
of our results
\citep[e.g.][]{hamill_social_2009}.
Furthermore, as we pointed earlier, our analysis was based on the static
view of homophily which does not differentiate between value/choice and
status/induced homophily as well as effects of selection and influence.
Thus, our results say little about the extent to which the typical properties
of social networks are shaped by any of these more specific social forces.
More dynamical analyses, both empirical and formal/analytical, are needed to
answer this question (some good examples can be found in:
\cite{centola_homophily_2007,anagnostopoulos_influence_2008,kossinets_origins_2009,carletti_emerging_2011,centola_choosing_2015}).
However, there are also some partial results indicating that homophily-based
selection may be more prevalent than influence
\citep{anagnostopoulos_influence_2008}.
If this is the case, then it is more justified to interpret our results
in terms of causal effects of homophily on the structure of social networks.

\section{Conclusion}

In this paper we studied the extent to which the homophily principle
\citep{mcpherson_birds_2001}
is sufficient to explain the typical structural properties of social networks
such as sparsity, high clustering, positive degree assortativity,
short average path lengths (small-worldness) and right-skewed
but not scale-free distributions.
To do so we conducted simulation studies based on the Social Distance Attachment
(SDA) model introduced by
\citet{boguna_models_2004},
which is a type of a random geometric graph (RGG)
\citep{dall_random_2002}.
We derived the form of the model from first principles based on the direct
correspondence between the homophily principle and RGGs
and showed that homophily is indeed sufficient to reproduce most of the typical
properties of social networks, but it does not imply the small-world property.
However, we also showed that even a small amount of random edge rewiring,
as done by \citet{watts_collective_1998},
is enough to guarantee it without destroying the structure induced by homophily.

We also extended SDA model and combined it with the configuration model
\citep{newman_random_2001}
in order to study homophilic networks with arbitrary degree distributions.
We used it to further examine the robustness of the effects of homophily.
We found that clustering is relatively independent of degree distribution
constraints, while degree assortativity depends strongly on the degree
distribution and only weakly on the strength of homophily.

Our results complement existing literature on linkages between social structure,
homophily and social networks by showing that under very general assumptions
homophily leads to network structures that are characteristic for many
observed social networks. This suggests that it should be considered one of
the important generative processes for social networks. Moreover,
based on our results we proposed a simple rule of thumb for distinguishing
between networks that are likely and unlikely to be shaped by homophily.

Last but not least, our results are of use also from a purely practical perspective,
since we provide researchers with two graph models that can be used to
efficiently simulate networks with a wide range of structural features typically
observed in real-world social networks.

\endparano


\section{Acknowledgements}

The authors thank Tomasz Zarycki for general discussions and the encouragement
and Ivan Voitalov for the advice on detection and classification of power law
tails as well as long discussions about random geometric graphs.
We~also express our gratitude to the anonymous reviewers for very helpful
comments. \\
\textbf{Author contributions:} S.T. and A.N. conceived the project
and planned the structure of the paper. S.T. wrote the paper, implemented
the models, run the simulations and performed the analyses. \\
\textbf{Funding.} A.N. acknowledges the support of a grant from
Polish National Science Center (DEC-2011/02/A/HS6/00231).


\newpage

\section[A]{Appendix A: example social spaces and network realizations (SDA)}
\label{app:sda}
\begin{figure}[h]
    \centering
    \includegraphics[width=.8\textwidth]{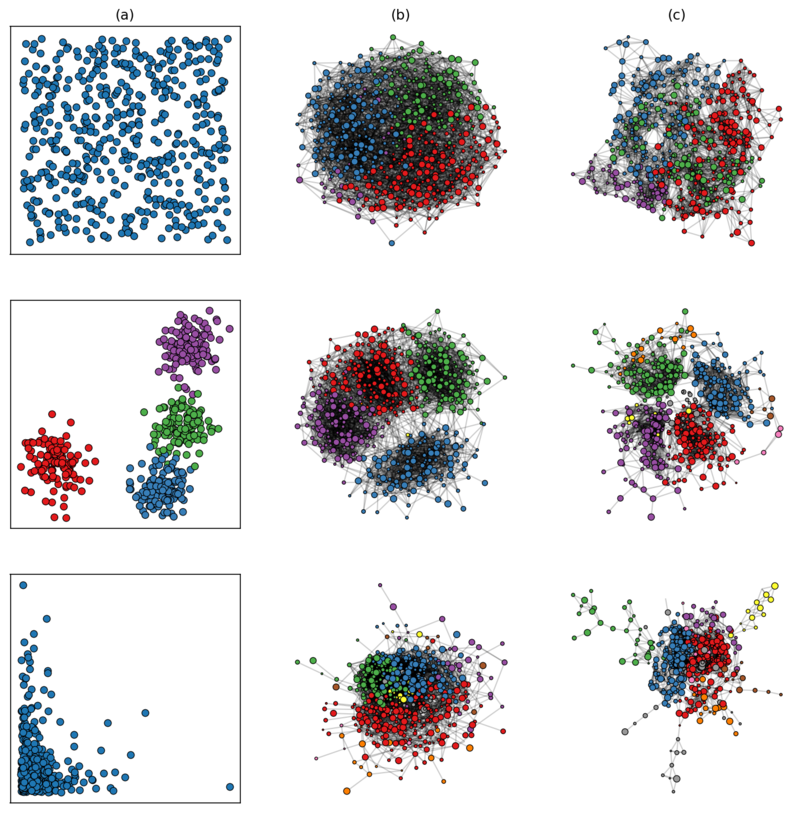}
    \caption{
        Example realizations of social space ($\mathsf{N = 500, m = 2}$)
        and corresponding networks. Middle graphs have $\mathsf{\alpha = 2}$ and
        rightmost have $\mathsf{\alpha = \infty}$. Kamada-Kawaii algorithm was used
        for network layout and nodes' colors denote communities as detected
        with the greedy modularity optimization
        \citep{clauset_finding_2004}.
    }
\end{figure}

\section[B]{Appendix B: SDC algorithm pseudo-code}
\label{app:sdc-algorithm}

Let $k_1, k_2, \ldots, k_N$ be a degree sequence for $N$ nodes
(must sum to an even number), $\mat{P}_N = (p_{ij}) \in [0, 1]$
be a $N$-by-$N$ connection probabilities matrix derived from SDA model
and $\mat{A}_N = (a_{ij}) = 0$ be an empty adjacency matrix filled with zeros.

\begin{enumerate*}
    \item Set $p_{\text{malformed}} \in (0, 1)$ to some preferably very small, but
    non-zero value (i.e. $10^{-9}$).
    \item For $i, j = 1, \ldots, N$:
    \begin{itemize*}
        \item If $p_{ij} = 0$: set $p_{ij} = p_{\text{malformed}}$.
    \end{itemize*}
    \item For $i = 1, \ldots, N$:
    \begin{itemize*}
        \item If $k_i = 0$: set $p_{ij} = 0$ for $j = 1, \ldots, N$.
    \end{itemize*}
    \item Select a node $n_i$ at random with selection probabilities
    $p(n_i) = \sum_{k = 1}^N p_{ik} / \sum_{j=1}^N\sum_{k=1}^N p_{jk}$
    and decrease its number of stubs $k_i$ by one.
    \item Select a node $n_j$ at random with selection probabilities
    $p(n_j) = p_{ij} / \sum_{k=1}^N p_{ik}$
    and decrease its number of stubs $k_j$ by one.
    \item Set $a_{ij} = a_{ij} + 1$.
    \item If $a_{ij} \geq 1$: set $p_{ij} = p_{\text{malformed}}$.
    \item Repeat steps 4-8 until $k_i = 0$ for all $i = 1, \ldots, N$.
\end{enumerate*}

Note that steps 5 and 6 ensure that at each iteration every available edge
is selected with probability proportional to its current $p_{ij}$. Morever,
the fact that self-loops and multiple edges may be created means that degree
sequences do not have to meet the formal realizability conditions
\citep{hakimi_realizability_1962}.

The algorithm is implemented in Python in the method \texttt{conf\_model}
of the \texttt{SDA} class in \texttt{sdnet/sda.py}.

\section[C]{Appendix C: simulating power law distributed degree sequences}
\label{app:simulate-pl}

Simulating power law distributed degree sequences for networks
of given size and with given $\E[k]$ is not a trivial task.
A standard generator of pseudorandom numbers with, for instance,
Pareto distribution will not do, because it may generate numbers bigger than
the size of a network and does not guarantee that a sequence will sum to
an even number. Moreover, in this case it is absolutely not clear as to
what particular value of the characteristic exponent $\gamma$ should be
chosen. To solve all these problems we simulated degree sequences via
the standard preferential attachment (PA) process
\citep{barabasi_emergence_1999}
with new nodes establishing $m = [\frac{1}{2}\E[k]]$ edges. This fixes
the average degree exactly to the desired value and determines
proper $\gamma$. Also PA process will of course never yield values larger than
the system size. However, it sets an artificial
lower bound on node degrees at $m$. To solve this, we also add uniform
integer noise in the range $[-m, m]$ and cap high values at $N-1$
(the maximum number of edges a node can have). \\

\begin{table}[h]
\centering
\begin{tabular}{cccccc}
    \toprule
    & \multicolumn{4}{c}{Power law class} & \\
    N    & NPL   & HPL  & PL  & DSM  & PL or DSM \\
    \midrule
    1000 &       & 36   & 11   & 3 & 14 \\
    2000 &       & 28   & 15   & 7 & 22 \\
    4000 &       & 27   & 20   & 3 & 23 \\
    8000 &       & 25    & 17  & 8 & 25 \\
    \bottomrule
\end{tabular}
\caption{
    Tail index classification for 200 simulated power law degree sequences
    of different sizes (50 sequences per size). The columns refer to the
    classes proposed by
    \citet{voitalov_scale-free_2019}:
    not power law (NPL), hardly power law (HPL), power law (PL),
    power law with divergent (infinite) second moment (DSM).
}
\label{tab:app:simulate_pl}
\end{table}

We tested validity of this approach using the methods for classifying tail
behavior developed by
\citet{voitalov_scale-free_2019}.
Table~\ref{tab:app:simulate_pl} shows the results. Degree sequences for
higher values of $N$ tend to be classified as PL or DSM more often.
Moreover, all degree sequences have been classified at least as HPL.
This, together with the fact that PA process yields power laws
only asymptotically, indicates that our method is reasonably effective.

\newpage
\section[D]{Appendix D: example degree sequences and network realizations (SDC)}
\label{app:sdc}
\begin{figure}[h]
    \centering
    \includegraphics[width=.8\textwidth]{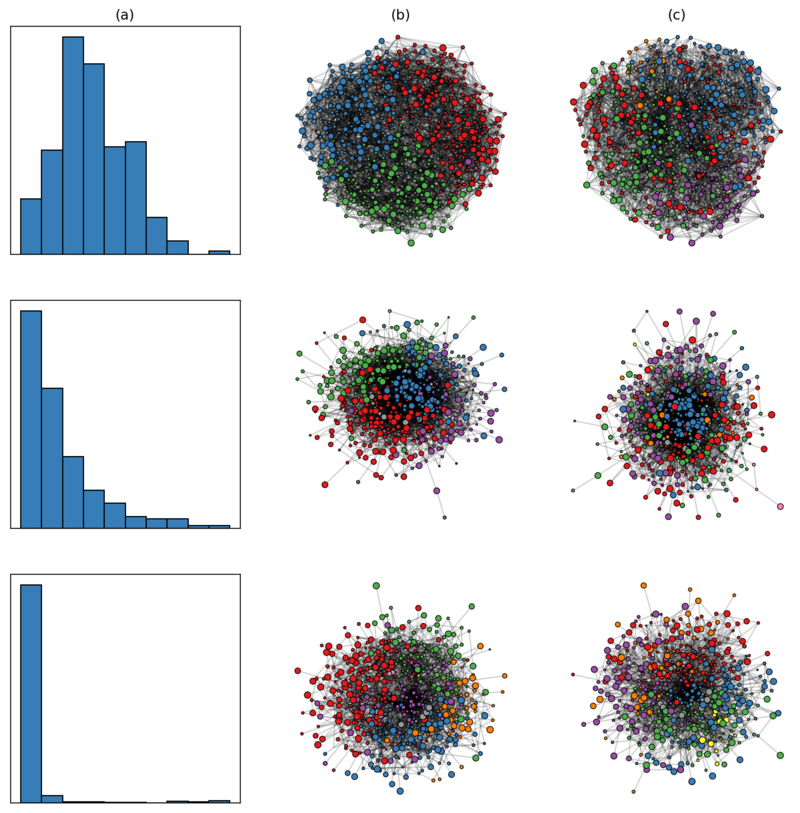}
    \caption{
        Example realizations of degree sequences and corresponding graphs
        ($\mathsf{N = 500, m = 2}$). Middle column present networks with low
        homophily ($\mathsf{\alpha = 2}$) and the right column shows networks with
        high homophily ($\mathsf{\alpha = \infty}$). Kamada-Kawaii algorithm
        was used for network layout. Nodes' colors denote communities detected
        with the greedy modularity optimization
        \cite{clauset_finding_2004}.
    }
\end{figure}





\bibliographystyle{jasss}
\bibliography{sda-paper-jasss} 


\end{document}